\documentclass[fleqn,twoside]{article}
\usepackage{cite,amsfonts}

\newcommand{\bls}[1]{\renewcommand{\baselinestretch}{#1}}
\def\noi{\noindent}
\bls{0.97}

\newcommand{\Title}[1]{\noi {\uppercase{\Large #1}} \\}
\newcommand{\Author}[2]{\noi{\large\bf #1}\\[2ex]\noindent{\it #2}\\}
\newcommand{\Abstract}[1]{\vskip 2mm \begin{center}
        \parbox{16.4cm}{\small\noi #1} \end{center}\medskip}

\def\nq{\hspace*{-1em}}
\def\nqq{\hspace*{-2em}}
\def\nhq{\hspace*{-0.5em}}

\def\cm{\hspace*{1cm}}
\def\inch{\hspace*{1in}}
\def\wide{\mbox{$\dst\vphantom{\int}$}}

\def\Jl#1#2{{\it #1\/} {\bf #2},\ }
\def\PLB#1 {\Jl{Phys. Lett.}{#1B}}
\def\PRL#1 {\Jl{Phys. Rev. Lett.}{#1}}
\def\al{&\nhq}
\def\lal{&&\nqq {}}
\def\eq{Eq.\,}
\def\eqs{Eqs.\,}
\def\beq{\begin{equation}}
\def\eeq{\end{equation}}
\def\bear{\begin{eqnarray}}
\def\bearr{\begin{eqnarray} \lal}
\def\ear{\end{eqnarray}}
\newcommand{\bd}{\begin{displaymath}}
\newcommand{\ed}{\end{displaymath}}
\def\earn{\nonumber \end{eqnarray}}
\def\nn{\nonumber\\ {}}

\def\nnn{\nonumber\\ \lal }

\def\yy{\\[5pt] {}}
\def\yyy{\\[5pt] \lal }
\def\eql{\al =\al}
\def\Sp{\mathop{\rm Sp}}

\def\dst{\displaystyle}

\def\e{{\,\rm e}}
\def\d{\partial}
\def\const{{\rm const}}
\begin{document}

\twocolumn[ \Title {A conformally invariant generalization of string
theory\yy
  to higher-dimensional objects.\yy Hierarchy of coupling constants}

\Author{F.Sh. Zaripov}
  {Department of Mathematics, Kazan State Humanitarian and Pedagogical University, 1 Mezhlauk St.,
  Kazan 420021, Russia; e-mail: farhat@kazan-spu.ru}

 \Abstract {We suggest a conformally
invariant generalization of string theory toi
 higher-dimensional objects. As such a model, we consider a conformally
 invariant $\sigma$ model. For this theory, the Hamiltonian formalism is
 constructed, and the full set of constraints is found. The equations
 obtained are studied under a fixed gauge. It is shown that special cases
 of the model are string theory and Einstein's theory of gravity.
 Cosmological application of the suggested theory are studied. It is shown
 that Friedmann-like models can be described in this framework. Our models
 make it possible to interpret the Universe evolution as evolution of
 three-dimensional objects embedded in a higher-dimensional flat space-time.}

] 

\section{A conformally invariant generalization of string theory to
    higher-dimensional objects}

  The remarkable achievements of string and superstring theories are well
  known: evaluation of the space-time dimension, fixing a particular gauge
  group, inclusion of gravity into a unified scheme etc. \cite {grin}.
  These achievements stimulate an interest in studies of geometric objects
  of higher dimension, such as membranes or p-branes. It is known, however,
  that \cite{ket} in standard membrane theories the absence of conformal
  invariance precludes the usage of string-theoretical methods. For
  instance, the requirement that conformal invariance should be preserved at
  the quantum level leads, in string theory, to fixing the space-time
  dimension \cite{brn}. There are also other arguments \cite{vit} in favour
  of the requirement that a physical field theory should be conformally
  invariant, at least at the classical level.

  On this basis, we have previouly suggested a conformally invariant
  generalization of string theory to higher-dimensional objects \cite{zar}.
  This paper, aimed at further realization of this approach, is devoted to
  obtaining and investigation of Hamiltonian equations and constraint
  equations of the theory under consideration. This idea was originally
  suggested as a quantum theory by analogy with string theory. However, a
  further analysis has shown the necessity of an initial classical analysis
  of this theory. It has turned out that even the classical level of the
  theory contains results of interest related to gravitation theory and
  p-brane theory. The action that serves as a basis for the suggested
  theory, being a generalization of string and p-brane theory, is
  simultaneously a certain generalization of Einstein's general relativity.
  We suggest that general relativity should be considered as a special case
  of a conformally invariant sigma model, appearing as a result of conformal
  symmetry violation.

  This paper is devoted to foundation and analysis of the above ideas in the
  classical case.

  The recent development of multidimensional theories have been, to a large
  extent, related to the so-called branes. In this theory [6--9], the
  observable Universe is considered as a surface (brane) embedded in a
  higher-dimensional space-time. It is hoped that this approach can lead to
  a success in solving the fundamental problem of the hierarchy of physical
  coupling constants and the cosmological constant problem. The hierarchy
  problem lies in the existence of a huge difference between the elecroweak
  energy scale of about 1 TeV and the gravitational energy scale of the
  order of $10^{19}$ GeV. Besides, the energy density related to the
  cosmological constant should be about 120 orders of magnitude smaller than
  the possible energy density values for the known models of quantum theory
  of the weak and strong interactions.

  In the theory suggested, the gravitational constant is related to the
  dynamic characteristics of the model, and it is obtained in
  multidimensional space-time due to localization of solutions to nonlinear
  equations, by analogy with the Higgs effect in gauge field theory.

\subsection{The Lagrangian approach}

  The action for a membrane (or p-brane) does not admit conformal
  transformations, and these models do not possess a natural candidate for
  the role of an anomalous symmetry like conformal symmetry in string
  theory. To circumvent this difficulty without abandoning the
  string-theoretical ideology, we suggest the following generalization of
  string theory:
\bearr\label{ns1}
    S = \frac{1}{w} \int \biggl\{ - \frac{1}{2}
  (\nabla_{\nu} X,\nabla^{\nu} X ) + {\tilde \xi} \stackrel{n}{R}(X,X)
\nnn \inch
    + \Lambda (X,X)^{\rho} \biggr\} \sqrt{- g} \, \hat d^{p+1} \sigma,
\ear
  where we use the notations:
\bear
    (X,X) \eql X^{A}X^{B}\eta_{AB},
\nn
    (\nabla_{\nu}{X},\nabla^{\nu}{X}) \eql
    \nabla_{\nu}X^{A}\nabla_{\mu}X^{B}g^{\nu\mu}\eta_{AB},
\nn
   \rho \eql (p+1)/(p-1).
\earn
  In the action (\ref{ns1}), the functions $X^{A}=X^{A}(\sigma^{\mu})$, with
  $A,\,B = 1,\ 2, \ldots , D;\quad \mu,\nu=0,1,\ldots,p$, map the
  $n = p+1$-dimensional manifold $\Pi$, described by the metric
  $g_{\mu\nu}$, into $D$-dimensional space-time $M$ with the metric
  $\eta_{AB}$, where the space $M$ is determined by the Minkowski metric
  with the signature $(-,+,\ldots,+)$. However, it turns out that, in a
  detailed study, it is more convenient to leave the signature of $M$
  arbitrary. The flat space signature is here understood as the set of signs
  of the elements along the main diagonal ($+1$ and $-1$) of the
  metric matrix. The quantity  $\stackrel{n}{R}$ is the scalar
  curvature of the manifold $\Pi$, the operator $\nabla_{\nu}$ means a
  covariant derivative in the manifold $\Pi$, where the Christoffel symbols
  are connected with the metric in the standard manner. We will
  assume that the space $\Pi$ is parametrized by the coordinates
  $\sigma^{\mu}$, where $\sigma^{0}= t$ is the temporal coordinate while the
  components $\sigma^i$ $(i=1,2,\ldots,p)$ describe a certain
  $p$-dimensional object, to be designated as $\Gamma$. The quantities $w$,
  ${\tilde\xi}$ and $\Lambda$ are constants. The models like that with the
  action (\ref{ns1}) are also often called nonlinear $\sigma$ models.

  The action (\ref{ns1}) is conformally invariant if
\beq\label{2}
    {\tilde \xi} = \xi \equiv - \frac{p-1}{8p}.
\eeq
  This invariance is expressed in the fact that the equations obtained by
  varying the action (\ref{ns1}) with respect to the fields $\hat g$ and
  $\hat X$ are invariant under the local Weyl scale changes
\beq \label{cp}
    g_{\mu\nu} \mapsto  \e^{2\phi} g_{\mu\nu}, \qquad
    X^{A} \mapsto  \e^{4\xi p\phi}X^{A},
\eeq
  for an arbitraryf $\phi = \phi(\sigma^{\mu})$.

  After varying the action (\ref{ns1}), the field equations for
  $\hat X$ and $\hat g$ have the following form:
\bearr \label{Y}
    Y^{A}\equiv \Box X^{A} + 2 \xi \stackrel{n}{R} X^{A} + 2 \Lambda
        \rho (X,X)^{\rho - 1} X^{A} = 0,
\nnn
\\ \lal \label{T}
    T_{\alpha\beta} \equiv T^{1}_{\alpha\beta} + \frac{2\xi}{w}[ -
 \stackrel{n}{R}_{\alpha\beta} + \frac{1}{2} \stackrel{n}{R} g_{\alpha\beta}
\nnn \cm
    + \nabla_{\alpha} \nabla_{\beta} - g_{\alpha\beta} \Box ](XX) = 0,
\ear
  where
\beq \label{T1}
    T^{1}_{\alpha \beta}= \frac{1}{w} [(\nabla_{\alpha}X,
        \nabla_{\beta}X) + L_1 g_{\alpha\beta} ]
\eeq
  are the terms appearing due to variation of the Lagrangian density
\beq \label{L_1}
    L_1 = - \frac{1}{2w} g^{\mu\nu} (\nabla_{\mu} X, \nabla_{\nu} X) +
        \frac{1}{w} \Lambda (X, X)^{\rho}.
\eeq

  If the action is supplemented by Lagrange functions of other matter
  fields, then \eq (\ref{T}) is replaced by the equation
\beq \label{T3}
    T_{\alpha\beta} + T^{e}_{\alpha\beta}=0,
\eeq
  where $T^{e}_{\alpha\beta}$ is the energy-momentum tensor of the other
  fields. In case $(X,X) = \const$, as follows from (\ref{T3}) and
  (\ref{T}), the equations are similar to Einstein's, with the canonical
  energy-momentum tensor $T^{1}_{\alpha \beta}$ and the effective
  gravitational constant
\beq \label{kap}
     G_e = -\frac {w_0} {16\pi\xi(X, X)}, \qquad w= \frac {w_0}{16\pi}.
\eeq                                                 

  Let us point out the important fact that for strings ($p = 1$) the general
  solution to \eqs (\ref{T}) has the form
\beq \label{gtti}                                    
    Bg_{\mu\nu} =
    (\nabla_{\mu} X, \nabla_{\nu} X), \qquad \mu, \nu = \overline{0,p},
\eeq
  where $B$ is an arbitrary function. Thus the original metric $g_{\mu \nu}$
  is connected by a conformal transformation with the induced metric
  $(\nabla_{\mu} X, \nabla_{\nu} X) $. Unfortunately, in the general case
  $p  > 1 $, for \eqs (\ref{Y})--(\ref {T}), the solution (\ref {gtti})
  is not a general solution. The problem of connection between the metric
  of the manifold $\Pi_g$ with the metric induced by the solutions
  $X^A = X^A (\sigma^{\mu})$, as well as that of a physical interpretation
  of this connection, have not been solved for an arbitrary dimension.

  In what follows, we will consider some special solutions to \eqs
  (\ref {Y})--(\ref {T3}), being of interest for physics.

\subsection{Hamiltonian formalism}

  To pass over to the Hamiltonian formalism, we make, in the action
  (\ref{ns1}), a $(p + 1)$-partition. Employing the results of
  Refs.\,\cite{tor,tut}, we introduce the parameters $N$ and $N_{i}$, the
  ``lapse'' or ``shift'' functions, and the metric functions of the
  $p$-dimensional geometry $h_{i j}$, where $i = \overline{1,p}$:
\bearr
    g_{00} = N_{s} N^{s} - N^2,\cm   g_{0 i} = N_{i},
\nnn
    g_{i j}= h_{i j},  \cm\cm \sqrt{-g} = N\sqrt{h}.
\earn
  Then, taking into account the results of Ref.\,\cite{tor}, we can present
  the scalar curvature in the form
\bearr\label{RN}
    \stackrel{n}{R} = R - (\Sp\hat K)^2 + \Sp(\hat K^2)
\nnn \cm
    - \frac{2}{N\sqrt{h}} \d_{\alpha}[N\sqrt{h}
    (n^\alpha \Sp\hat K + \nabla_{\beta} n^{\alpha})].
\ear
  Here $R$ is the scalar curvature calculated for the metric $h_{ij}$,\quad
  $a^{\alpha} = \nabla_{\beta} n^{\alpha} n^\beta$ is the
  $(p+1)$-dimensional acceleration of an observer moving along a timelike
  normal $\vec n$ to consecutive sections. The space-time $\Pi$ is assumed
  to be foliated into a one-parameter family of spacelike hypersurfaces
  with the parameter $t$. The quantity $\hat K$ is the extrinsic curvature
  tensor of the spacelike sections:
\bearr \label{naa}                                                  
    n_{\alpha} = \left\{- N, 0\right\},\qquad (\vec n \cdot \vec n)= - 1,
\\ \lal \label{spp}
    \Sp\hat K = h^{i j} K_{ij},\qquad \Sp\hat K^2 = K_{ij} K^{ij},
\nnn
    K_{ij} = \frac{1}{2N}(D_i N_J + D_j N_i - \d_t h_{ij}),     
\ear
  where $D_i$ is a covariant derivative calculated with the metric $h_{ij}$.

  Let us now pass over to a description in terms of the phase-space
  variables, i.e., the generalized coordinates and momenta:
\[   \nq\,
    \{ q_I \} = \{ X^A, h_{lk}, N, N_i \},\quad
    \{ p_I \} = \{ P_A, \Pi^{lk}, p_N, p_{N_i} \},
\]
  where
\beq\label{PI}
    p_I = \frac{\delta L}{\delta (\d_t q_I)} ,
\eeq
  where $L$ is the Lagrangian corresponding to the action (\ref{ns1}).
  We obtain as a result:
\bearr
    P = \frac{\sqrt{h}}{N w}[\dot X - N^i D_i X + 4 \xi X N \Sp {\hat K}],
\nnn
    \Pi^{lk} = \xi \frac{\sqrt{h}}{w}
            \biggl[(X,X)(\Sp {\hat K}\cdot h^{lk} -K^{lk})
\nnn \cm\ \
    - \frac{1}{N} h^{lk}( \d_t (X,X) -  N^i D_i (X,X))\biggr],
\ear
  the remaining momenta are zero. In the conformal transformations
  (\ref{cp}), the phase variables are transformed as follows:
\bearr
    h_{lk} \mapsto \e^{2 \phi} h_{lk} \ ,\quad X \mapsto
    \e^{4\xi p \phi} X \ ,\quad N \mapsto \e^{\phi} N ,
\nnn   \label{pl}
    N_i \mapsto \e^{\phi} N_i,\qquad
    P \mapsto  \e^{-4\xi p \phi}P ,\quad\
    \Pi^{lk}\mapsto  \e^{-2\phi} \Pi^{lk}.
\nnn
\ear

  As follows from (\ref{pl}), there is a constraint between $P$ and
  $\Pi^{lk}$. This constraint may be written as
\beq \label{M1}
    M \equiv 2 \Sp\hat {\Pi}  + 4 \xi p (P,X) = 0.
\eeq
  Integrating by parts and rejecting terms with a full divergence,
  one can write the action (1) in terms of the canonical variables as
\bearr \label{S}
    S = \int [ \dot h_{lk} \Pi^{lk} + (P,\dot X) - N H_0 - N_i H^i
\nnn \inch
    - \lambda_M  M - \d_i Q^i ] \hat d^{p+1} \sigma ,   
\ear
  where
\bearr
    H_0 \equiv \frac{w}{\xi \sqrt{h} (X,X)}[ \Sp(\hat \Pi \hat \Pi) -
        \frac{1}{p} (SP \hat \Pi)^2] + \frac{w}{2\sqrt{h}} P^2
\nnn \cm    \label{H0}
     + \frac{\sqrt{h}}{w} \biggl[ - \xi (X, X) R + \frac{1}{2}(D_s X,D^s X)
\nnn \inch\ \
     +  2 \xi \Delta(X,X) -  \Lambda (X, X)^{\rho}\biggr],     
\yyy \label{Hl}
        H^l \equiv ( P, D^l X ) - 2 D_s \Pi^{sl} ,        
\yyy\label{Qi}                                                    
    Q^i = \frac{2\sqrt{h} \xi}{w} [ (X, X) D^i N - N D^i (X, X) ]
\nnn \cm
    + 2 N_k \Pi^{ki} + \frac{N^i}{4} \biggl[\frac{1}{\xi p} \Sp\hat \Pi +
        \frac{1}{w N} E \biggr].
\ear
  Here we have used the notations
\beq \label{E}
    E = \d_t (X, X) - N^s D_s (X,X) ,
\eeq
  $\Delta = h^{lk} D_l D_k$ \ and \ $(D_s X,D^s X) = (D_i X, D_j X) h^{ij}$.
  The function $\lambda_M$ is arbitrary. This is related to the
  impossibility of resolving the velocity $\lambda_E = \d_t (X,X)$
  in terms of the momenta. It can be shown that
\[
   p_E \equiv \frac{\delta L}{\delta \lambda_E} = \frac{M}{ 8 \xi p(X,X)}.
\]
  The constraint (\ref{M1}) has appeared because of the invariance of the
  theory with respect to(\ref{pl}). To take this fact into account
  explicitly, let us transform the integrand in (\ref{S}) according to
  (\ref{pl}) for $\phi = \psi$ and introduce the field $\psi$. Then the
  expression (\ref{S}) takes the form
\bearr \label{S1}
    S = \int [ \dot h_{lk} \Pi^{lk} + (P,\dot X) - N H_0 - N_i H^i
\nnn \cm
    - (\lambda_M - \dot \psi + N^s D_s \psi) M -
            \d_i \tilde Q^i ] \hat d^{p+1}\sigma.   
\ear
  That is, we could use, instead of $\lambda_{M}$, the field $\psi (\sigma)$.
  The corresponding momentum is $p_{\psi} = M$. The divergence term in
  \eq (\ref{S1}) does not affect the equations of motion but affects the
  boundary conditions. After the transformations (\ref{pl}), the quantities
  $Q^i$ turn into $\tilde Q^i$, where
\bearr \label{Q1}
    \tilde Q^i =  Q^i + (X, X) \frac{\sqrt{h}}{w}
\nnn \cm
    \times \biggl[ 2 \xi p N D^i \psi  +
        \frac{ \xi p N^i}{N} (\dot {\psi} - N_s D_{\psi} ) \biggr].
\ear
  The latter expression implies that, taking into account the boundary
  effects, we shall obtain certain boundary conditions applied to the
  function $\psi$, violating the invariance of the theory with respect to
  (\ref{pl}). It is probably reasonable to omit the divergence term
  (\ref{Q1}) from the action, replacing the original Lagrangian density
  with $L + \d_i Q^i$. An argument in favour of such a replacement is that
  for $(X, X) = \const$ and $\ X^i = \const$, the action acquires the
  Einstein form. In the construction of the Hamiltonian formalism for
  Einstein's theory, such terms are omitted \cite{tut}.

  The conditions that the primary constraints are conserved in time,
\beq \label{Phi1}
    \Phi^{(1)}_{I}: \ \ p_{\mu} \equiv \{ p_N, p_{N_{i}}\} = 0,
    \qquad   p_{\psi} - M = 0,                                 
\eeq
  with the Hamiltonian constructed in the standard way \cite{tut},
\beq \label{H}
    H = \d_i Q^i + N H_0 + N_s H^s + \lambda_0 M               
\eeq
  and the extended Hamiltonian
\[
    H^1  = H + \lambda_I \Phi^{(1)}_{I},
\]
  do not allow one to determine the functions $\lambda_{I}$,
  but there emerge secondary constraints:
\beq \label{Phi2}                                                   
    \Phi^{(2)}_I : \ \
        H^{v} \equiv \left\{ \  H_0, H^l, M \ \right\} = 0.
\eeq

  Consider the conservation conditions for the constraints (\ref{Phi2}). To
  do so, it is necessary to calculate the Poisson brackets:
\[
    [ \Phi_K, \Phi_J ] \equiv \ \frac{\delta \Phi_K}{\delta q_I}
 \frac{\delta \Phi_J}{\delta p^I} \ - \ \frac{\delta \Phi_K}{\delta p^I}
        \frac{\delta \Phi_J}{\delta q_I} .
\]
  After cumbersome calculations, it can be shown that if the appearing
  divergence terms, leading to surface integrals, vanish, then the
  constraint conservation conditions do not allow determining the functions
  $\lambda_I$ and do not lead to new constraints. All constraints are
  thus first-class constraints.

  The equations of motion $\dot q = [q, H]$ have the following form:
\bearr                     \label{x1}
    \dot X = \frac{Nw}{\sqrt{h}} P + N^s D_s X + 4\xi p\lambda_M X,
\yyy \label{x2}
     \dot h_{lk} = \frac{2wN}{\xi \sqrt{h}(X, X)}
    \biggl[ \Pi_{lk} - \frac{1}{p} (\Sp \Pi) h_{lk} \biggr]
\nnn \inch
    + D_{(l}N_{k)} + 2\lambda_M h_{lk},                      
\yyy
     \dot P = \frac{2wN}{\xi \sqrt{h} (X,X)^2}
        \biggl[ \Sp(\hat {\Pi}\hat {\Pi}) -\frac{1}{p}(\Sp\hat \Pi)^2 \biggr]X
\nnn
    + \frac{2\xi \sqrt{h}}{w}[ R N - 2\Delta N] X          \label{x3}
    + \frac{\sqrt{h}}{w}( N \Delta X + D^s N D_s X )
\nnn
    + \frac{2\rho N\sqrt{h}}{w}\Lambda (X, X)^{\rho - 1} X
            + D_s(N^s P) - 4 \xi p \lambda_M P,     
\nnn
\\ \lal
     \dot \Pi^{lk} = - \frac{2wN}{\xi \sqrt{h}(X, X)}
        \biggl[\Pi^{l}_{m} \Pi^{mk} -\frac{1}{p}\Sp(\hat \Pi) \Pi^{lk}\biggr]
\nnn
    + \frac{N}{\sqrt{h}}h^{lk}\left\{ \frac{w}{\xi (X, X)}[\Sp(\hat \Pi
    \hat \Pi) - \frac{1}{p}(\Sp\hat \Pi)^2] + \frac{w}{2} P^2 \right\}
\nnn
    + \frac{\xi \sqrt{h}}{w} [ \ N D^l D^k (X, X) - N R^{lk} +
            (X, X) D^l D^k N
\nnn \cm
    - h^{lk} \big( \, (X,X)\Delta N + D_s(X, X) D^s N \,\big ) \ ]
\nnn  \label{plk}
    + \frac{N\sqrt{h}}{2w}(D^l X, D^k X) -
        c^{lk} + \frac{1}{2} P N^{(l} D^{k)} X
\nnn \cm                                                            
    - 2\lambda_M \Pi^{lk} -\frac{Nw}{2\sqrt{h}(X,X)} h^{lk} H_0,
\ear
  where
\[
   c^{lk} = \sqrt{h} D_s\biggl(\frac{1}{\sqrt h} ( N^{(l}\Pi^{k)s}
    - N^s \Pi^{lk} \ )\biggr)
\]
  In what follows, we will put the constant $w$ equal to unity. If, instead
  of the indefinite coefficient $\lambda_M$, we introduce the field $ \psi$,
  we should make the following substitution in the equations of motion:
\beq
    \lambda_M = \lambda_0 - \dot \psi + \psi_s N^s,            
\eeq
  where $\lambda_0$ is an arbitrary function of the phase variables. This
  function may be chosen to be equal to zero, which simply re-defines the
  function $\psi$. Then, using the substitutions
\bearr
    h_{lk} = \e^{- 2\psi} \bar {h}_{lk} ,\qquad
    X = \e^{- 4\xi p \psi} \bar X,  \qquad N= \e^{- \psi} \bar N ,
\nnn   \label{pk}
    N_i = \e^{- \psi} \bar N_i,\qquad\ P = \e^{4\xi p \psi} \bar P,
        \qquad\ \Pi^{lk} = \e^{2\psi} \bar {\Pi}^{lk},      
\nnn
\ear
  one can exclude the field $\psi$ from \eqs (\ref{x1})--(\ref{plk})
  and pass over to the conformally invariant canonical variables
  $\{ \bar {q_I}, \bar {p^I} \}$, which is equivalent to putting
  $\lambda_M = 0$ in the equations. However, for studying different gauge
  conditions, it is more convenient to preserve the arbitrariness in
  choosing the function $\lambda$. To impose the canonical gauge, it is
  necessary to impose $2p+4$ supplementary conditions, according to the
  number of first-class constraints. We will consider as such constraints
  the class of additional conditions $\Phi_G$ of the form
\bearr
     N = \tilde N,\qquad N_l = \tilde N_l,\qquad \lambda  = 0,
\nnn \cm
     \chi_{\mu} = 0,\qquad F = 0,
\ear
  where $\chi_\mu$ are $p + 1$ functions of the phase variables $h_{lk}$ and
  $X^A$, while $F$ is a function of the phase variables $h_{lk}, \ X^A, \ P,
  \ \Pi^{lk}$. These functions are chosen in such a way that $\det [\hat
  \Phi, \hat \Phi] \not = 0$, where $\hat \Phi = \{\Phi, \Phi_G \}$.
  Let us denote $H^{v} = \{H_0, H^l,M\}$ and $\chi_u=\{\chi_\mu, F\}$, then,
  in the case under consideration,
\beq
    \det[\hat \Phi, \hat \Phi] = (\det [H^{v}, \chi_{u} ])^2
    \ ( [\lambda, p_{\psi}])^2 .
\eeq
  A gauge, related to a choice of the function $F$, violates the conformal
  symmetry and determines a ``representative'' from each class of
  conformally equivalent metrics. To reach comprehension of the different
  kinds of gauge conditions, let us consider some consequences of the
  constraint equations (\ref{Phi2}) and the equations of motion (\ref{x1}),
  (\ref{plk}). Let us define the conformally invariant tensor $\Theta_{lk}$
  which is traceless on the surface of the constraints:
\bearr
    \Theta_{lk} \equiv 2\xi \sqrt{h} N \biggl[ D_l D_k (X, X) -
            \frac{h_{lk}}{p} \Delta (X,X)
\nnn \inch
        - (X,X)( R_{lk} - \frac{h_{lk}}{p}R )\biggr]
\nnn   \label{Th}
    + 2\xi \sqrt{h} \biggl[ (X,X)( D_l D_k N -
                \frac{h_{lk}}{p} \Delta N ) \biggr]
\nnn \cm
    + N \sqrt{h} \biggl[ (D_l X, D_k X) -
                \frac{h_{lk}}{p} (D_s X, D^s X) \biggr]
\nnn \inch
        + \frac{h_{lk}}{p} H_0.                             
\ear
  Then, using \eqs (\ref{x1}) and (\ref{plk}) as well as the definitions
  (\ref{H0}), (\ref{Hl}) and (\ref{M1}), it is easy to prove the following
  identity:
\bearr \label{TH}
    \d_t \biggl[\Pi^l_k - \frac{\delta^l_k}{p} \Pi^s_s \biggr]
    - D_s\biggl[ N^s (\Pi^l_k - \frac{\delta^l_k}{p} \Pi^s_s )\biggr]
\nnn \cm
    =F^l_k + \frac{1}{2} \Theta^l_k + \frac{1}{2} N^{(l} H^{s)}
            h_{sk}-\frac{\delta^l_k}{p}N^{s} H_{s},      
\ear
  where
\[
    F^l_k = \Pi^{l}_{s}  D_{k} N^{s} -\Pi^{s}_{k} D_{s} N^{l}.
\]
  The latter equation is equivalent to \eq (\ref{plk}) provided the
  conditions (\ref{x1})--(\ref{x3}) hold.

\section{Induced gravity}

  Let us call the ``partial embedding'' condition the choice of the
  supplementary conditions $\Phi_G$ obtained from the requirement
\beq \label{gi}
    h_{lk} = B (D_l X, D_k X), \qquad  l,k = \overline{1,p},
\eeq
  where $B$ is a certain function. Thus the metric $h_{lk}$, entering into
  the original action, is connected with the induced metric
  $(\nabla_{l} X, \nabla_{k}X)$ by a conformal mapping.

  In the string case, the general solution to the constraing equations
  has the form
\beq \label{gj}
    Bg_{\mu \nu} = (\nabla_{\mu} X, \nabla_{\nu} X), \qquad
        \mu, \nu = \overline{0,p},
\eeq
  where $B$ is an arbitrary function. This solution follows from (\ref{Th})
  and (\ref{TH}) if one puts the momenta $\Pi^{lk}$ and the parameter $\xi$
  equal to zero. For an arbitrary dimension, \eq (\ref{gj}) (for $B=1$)
  determines the condition of full embedding of the mainfold $\Pi$ into
  the space-time $M$. This equation, written in the $p+1$-partition
  formalism, is equivalent to \eq (\ref{gi}) and the equations
\bear \label{kal}
    (\dot X, \dot X) \eql B(N_sN^s - N^2),            
\nn
    (D_l X, \dot X) \eql B N_l, \qquad B \equiv (D_l X, D^l X)/p =1.
\ear

  Thus we can consider two kinds of solutions corresponding to ``partial''
  or full embedding. In the first case, the validity of \eq (\ref{gi})
  (for $B=1$), as well as for strings, would permit one to interpret the
  fields $X^A$ as the conventional coordinates of a $d$-object $\Gamma$ in
  the space-time $M$. In other words, this means that, from each class
  $\Gamma$ of conformally equivalent manifolds, it is possible to choose at
  least one ``representative'' $\Gamma_0$, such that the functions $X^A$
  perform embedding of $\Gamma_0$ into the surrounding space $M$. Here,
  conformally equivalent manifolds are understood as manifolds whose metrics
  are connected with the reparametrization invariance and the conformal
  invariance (\ref{pl}).

  An invalidity of the relations (\ref{kal}), if (\ref{gi}) is valid,
  leads to some difficulties in the physical interpretation. If, by analogy
  with string theory, the space-time $M$ is considered as physical
  space-time, then the coincidence between the original metric $h_{lk}$ and
  the induced metric $(D_l X, D_k X)$ makes the theory transparent, making
  it possible to interpret the solutions $X^A = X^A (t, \sigma)$ as an
  embedding of a $p$-dimensional object $\Gamma$ into the physical
  space-time $M$. However, a non-coincidence between the ``lapse'' or
  ``shift'' functions of the original manifold $\Pi$ and the
  $p+1$-dimensional ``world history'' mainfold of the object $\Gamma$ poses
  a question on the physical meaning of the original functions $N$ and
  $N_i$. To answer this question, one can try to invoke the ideas of the
  Kaluza-Klein theory. We, however, put forward a conjecture according to
  which it is possible, at the expense of a choice of the corresponding
  reference frame and conformal gauge, and maybe also the dimension $D$,
  to achieve the validity of the conditions of full embedding of the whole
  manifold $\Pi$ into the space-time $M$. \eqs (\ref{Th}) and (\ref{TH})
  determine $\frac{p(p+1)}{2} - 1=k_p$ equations, while the number of
  arbitrary functions, determining the constraints $\Phi_G$, is equal to
  $2p + 4$. It is necessary to specify $p+2$ functions $N,N_l,\lambda_M$.
  Besides, according to the number of first-class conditions, we shoud
  impose $p+2$ supplementary conditions. One can try to impose the latter
  relations by requiring the $p+2$ conditions (\ref{kal}) to be valid. As
  follows from \eqs (\ref{Th}) and (\ref{TH}), to fulfil the ``partial
  embedding'' conditions (\ref{gi}), the following equations should hold:
\bearr
     \d_t \biggl[\Pi^l_k - \frac{\delta^l_k}{p} \Pi^s_s \biggr]
     - D_s\biggl[ N^s (\Pi^l_k - \frac{\delta^l_k}{p} \Pi^s_s )\biggr] -
     \frac{1}{2} F^l_k
\nnn \label{kali}
    = \xi \sqrt{h} \biggl\{ N \biggl [ D_l D_k (X, X) -
            \frac{h_{lk}}{p} \Delta (X,X)
\nnn \cm
    - (X,X)( R_{lk} - \frac{h_{lk}}{p} R )\biggr]              
\nnn \cm
    + \biggl[ (X,X)\biggl( D_l D_k N - \frac{h_{lk}}{p} \Delta N
            \biggr) \biggr] \biggr\}.
\ear
  If \eq (\ref{kali}) holds, then (\ref{gi}) follows from (\ref{Th}) and
  (\ref{TH}) The $p+2$ functions $N,\ N_l,\ \lambda_M$ should be chosen in
  such a way that, due to this choice, \eqs (\ref{kali}) hold.  The number
  of these functions for the dimensions $p=2$ and $p=3$ is 4 and 5,
  respectively, while the number of equations (\ref{kali}) (the number
  $k_p$) is 2 and 5, respectively. This simple counting of the degrees of
  freedom shows that, in the cases of interest $p = 2$ and $p = 3$, it is
  possible to choose a full embedding gauge.

  In the most general case, it can be proved that, to satisfy the full
  embedding conditions (\ref{gj}) (for $B=1$), it is necessary that
  the following equations hold:
\beq \label{ind}
    [\xi \stackrel{n}{R} + \rho (X, X)^{\rho-1}\Lambda] \d_{\mu}(X,X) = 0,
\eeq
  where $\mu = \overline{0, p}$. The simplest proof can be performed with
  the aid of the generally covariant equations (\ref{Y}). They are
  equivalent to the Hamiltonian equations (\ref{x1})--(\ref{x3}). Acting
  with the covariant derivative $\nabla_{\gamma}$ on the relation (\ref{gj})
  and contracting different pairs of indices, we obtain the equations
\bearr
   (\nabla_{\nu} X, \Box X) + (\nabla^{\mu} X, \nabla_{\mu} \nabla_{\nu} X)
            = \nabla_{\nu}B,
\yyy
    2 (\nabla^{\mu} X, \nabla_{\mu} \nabla_{\gamma} X)
                            =\nabla_{\gamma}B (p+1),
\ear
  where  $\mu, \nu, \gamma = \overline{0, p}$.

  From the latter equations combined with (\ref{Y}), we obtain \eq
  (\ref{ind}). In the derivation, we taking into account the covariant
  constancy of the metric tensor $g_{\mu \nu}$ and that $B=1$. Thus, as
  follows from (\ref{ind}), we can use two kinds of supplementary gauge
  conditions agreeing with the full embedding condition:
\bearr \label{oda1}
    (1) \quad F\equiv (X, X) - C =0, \quad C = \const;
\yyy \label{oda2}
    (2) \quad F\equiv \xi \stackrel{n}{R} + \rho(X,X)^{\rho-1}\Lambda = 0.
\ear
   As follows from (\ref{T})  and (\ref{oda1}), in the first case the
   constraint equations are similar to Einstein's equations with the
   canonical energy-momentum tensor (\ref{T1}) and the effective
   gravitational constant (\ref{kap}). For the second case, one cannot
   exclude solutions in which $G_e$ is variable and coordinate-dependent.

\subsection{Model theory}

   The equations obtained, like Einstein's equations, are strongly
   nonlinear and cannot be solved in a general form. However, the existing
   additional conformal symmetry simplifies the search for solutions of
   these equations. In this section, we simplify the equations obtained by
   restricting the class of metrics under consideration. Paying more
   attention to the dimension $n=4$, let us consider a model problem with
   the metric tensor $h_{lk}$ chosen in the form
\beq \label{mod}
    h_{lk} \ = \ b^2 \omega_{lk},
\eeq
  where $b^2 \ = \ b^2(t, \sigma)$ is an arbitrary function and $\omega_{lk}$
  is some fixed metric. It will be essential for what follows that the
  functions $\omega_{lk}$ are time-independent: $\dot \omega_{lk} = 0$.
  In the two-dimensional case, the following relations always hold:
\beq \label{rij}
     \stackrel{\omega}{R}_{lk} - \frac{\stackrel{\omega}{R}}{p}\omega_{lk}=0.
\eeq
  The index $\omega$ means that the corresponding quantities are calculated
  for the metric coefficients $\omega_{lk}$. For instance, $\omega_{lk}$
  may be chosen to be the metric of a constant-curvature space. Then,
\beq
    \stackrel{\omega}{R}_{lk} \ = k_0
            (p - 1)\omega_{lk} = -8\xi k_0 p \omega_{lk},
\eeq
  $k_0 = \{0, 1, -1\}$ for surfaces of zero, positive and negatice
  curvature, respectively.

  Here and henceforth, we leave the dimension arbitrary, considering
  simultaneously two-dimensional $\Gamma$ objects and objects of an
  arbitrary dimension, however, for the latter we restrict ourselves to
  spaces which are conformal to constant-curvature spaces.

\subsection{The ``conformal time'' gauge}

  Let us impose the following conditions on the ``lapse'' and ``shift''
  functions:
\beq     \label{n1}
        N^2 = b^2, \quad  N_i \ = \ 0.                  
\eeq
  After taking the trace of \eq (\ref{x2}), it follows:
\beq                                                            
     \lambda_M = \frac{\dot b}{b} = -
    \dot {\psi} \  \Longrightarrow \Pi_{lk} = \frac{1}{p}(\Sp \Pi)h_{lk}.
\eeq
  With the constraint $M=0$, we obtain
\beq
    \Pi_{lk} = -2\xi (P,X)h_{lk}.
\eeq

  Consider the gauge condition (\ref{oda1})
\beq \label{C}
    (X,X)- C \ = \ 0, \qquad  C = \const \not = 0.
\eeq

  Then, substituting  $X = g b^{4\xi p}$, from (\ref{x1}) and (\ref{x3})
  we obtain the equations for the field $g^A$ components:
\beq \label{gx}                                                 
    \ddot g - \stackrel{\omega}{\Delta} g -2\xi g
        \stackrel{\omega}{R} = 2\rho \Lambda C^{\rho-1} b^2g .
\eeq

  The constraint equations $H_0=0$ and $H_l=0$ may be brought to the form
\bearr \label{xxx}
    (\dot X, \dot X) = -(4\xi p \lambda_M)^2 C
\nnn \cm\cm
    + b^2 (2\xi C R + 2\Lambda C^{\rho} - Bd),        
\yyy                         \label{xx2}
    (\dot X,D_l X) = 16 \xi^2 p C D_l (\frac{\lambda_M}{N}).
\ear
   \eqs (\ref{kali}) are reduced to the following ones:
\bearr \label{rsii} \nq
    8 \xi p C \biggl(- D_l D_k \psi + D_l \psi D_k \psi -
        \frac{h_{lk}}{p}(- \Delta \psi+ D_s \psi D^s \psi ) \biggr)
\nnn \inch
    + C (\stackrel{\omega}{R_{lk}} -
        \frac{\omega_{lk}}{p}\stackrel{\omega}{R} ) = 0.
\ear
   Using (\ref{xxx}), (\ref{xx2}) and the consequence of \eq (\ref{C})
\[
    ( \dot X, \dot X) = - ( \ddot X, X),
\]
   we obtain an equation for finding the conformal factor:
\bearr \label{psii}
      \ddot \psi + 4\xi p \dot \psi^2 - \frac{1}{2p} b^2
    \big (R + 2 \Delta \psi
\nnn \inch
    - 2D_s \psi D^s \psi -2 (p+1) q \big ) = 0.
\ear

  The scalar curvature may be expressed in terms of the function $\psi$:
\beq\label{RR}
    R =\stackrel{\omega}{R} b^{-2} + 8 \xi p\, [ - 2
                \Delta \psi - (p-2)\psi_s  \psi^s \, ].
\eeq

  If one requires that the first and the second relations of the condition
  (\ref{kal}) hold, this leads to the equations
\bearr \label{goi12}
    - 8\xi p^2 \dot \psi^2 + b^2 (R + 16 \xi p q) = 0,      
\yyy       \label{goj1}
    D_l \dot \psi = - \dot \psi D_l \psi,                   
\ear
  where
\[
     q =\frac{1}{4\xi C} \biggl(B + \Lambda C^\rho \frac{1}{4\xi p}\biggr).
\]
  It can be shown that (\ref{psii}) is a differential consequence of
  (\ref{goi12}) and (\ref{goj1}). Then \eq (\ref{psii}) can be brought
  to the form
\beq
    p \ddot \psi + b^2( -2 q + \Delta \psi -                 
                    D_s \psi D^s \psi) = 0,
\eeq
  or, in terms of the metric $\omega$,
\beq \label{gpsi}
    p \ddot \psi + -2q b^2 +  \stackrel{\omega}{\Delta}\psi
        +8\xi pD_{\tilde s} \psi D^{\tilde s} \psi  = 0,   
\eeq
  Thus the function $\psi$ is found by solving \eqs (\ref{goi12}),
  (\ref{goj1}) and (\ref{rsii}).

  Then the functions $X^A$ are determined by \eq (\ref{gx}). If we write
  \eq (\ref{gx}) directly in terms of the variables $X^A$, we obtain a
  linear equation with respect to $X^A$. This equation, with (\ref{psii}),
  may be written as
\bearr \label{gxx}
    \ddot X - 8\xi p \dot \psi  \dot X  -\stackrel{\omega}{\Delta} X
    + 8 \xi p  D_{\tilde s} X D^{\tilde s} \psi - \frac{p+1}{C} b^2 X= 0.
\nnn
\ear
   Solutions of the latter equations should satisfy the remaining constraint
   equations which have the form
\bearr \label{xxxx}
    (\dot X, \dot X) = - b^2, \cm\ \   (X, X) = C,
\yyy \label{xx2x}
    (\dot X,D_l X) = 0.  \qquad (D_k X, D_l X) = h_{kl}.
\ear

  Let us present some special solutions to the equations obtained for the
  case of conformally flat manifolds. Let us first consider a flat
  $p$-dimensional model:
\[
    \stackrel{\omega}{R} =0.
\]
  Let the metric matrix $(\omega_{lk})$ be a unit matrix. A solution to \eqs
  (\ref{goi12}), (\ref{goj1}) and (\ref{rsii}) has the form
\beq \label{bab}
    b^{-2} \equiv  \e^{2\psi}=
    \biggl[ \frac{u_0}{2} ((r^2 - t^2) + mt +n_i \sigma^i + l_0)\biggr]^2,
\eeq
  where
\[
    r^2 = \sum_{i=1}^{p} (\sigma^{i})^2,
\]
  $u_0,\ m,\ n_i,\ l_0$ being integration constants satisfying the condition
  $m^2 +2 u_0 l_0 - n_i n^i = {2q}/{p}$. Here $t$ is the time coordinate,
  and the $p$-dimensional coordinates may be interpreted as the conventional
  Cartesian coordinates. It can be verified by a direct inspection that
  the functions $f_0=a_0 \dot \psi$ and  $f_l= a_l D_l \psi$ (where $a_0,
  a_l = \const$) are special solutions to \eq (\ref{gx}). Using this, let us
  build solutions which also satisfy \eqs (\ref{xxxx})--(\ref{xx2x}).
  Probably, there can be many such solutions. But we will here seek
  solutions with a minimal set of fields $X^A $. With this approach, we
  consider solutions which describe an embedding of the manifold $\Pi$ into
  the 5-dimensional space $M$. Calculations show that the solutions linear
  in the functions $f_0$ and $f_l$ satisfy \eqs (\ref{xxxx})--(\ref{xx2x})
  with the following values of the constants:
\bearr \label{cosm}
    q = \frac{p}{2C} \ \Longrightarrow \
            C^{\rho} = \frac{(p-1)(p+3)}{8 \Lambda},
\yyy \label{con1}
    a_0^2 = a_l^2 = \frac{1}{u_0^2}=C c_0,
\ear
  where $c_0=2 l_0/u_0$. Without losing generality in the solution
  (\ref{bab}), we put $m = n_i =0$, which simply corresponds to a parallel
  transport. Then the scale factor (\ref{bab}) is rewritten in the form
\beq   \label{bab1}
    b^{2} = \frac{4 a_0^2}{ (r^2 - t^2 + c_0)^2}.
\eeq
  ˆ§ (\ref{con1}) it follows that all solutions split into two types:
\bearr
    (1) \qquad  C > 0, \qquad c_0 > 0;
\nnn
    (2) \qquad  C < 0, \qquad c_0 < 0.
\earn
  To embed the manifold $\Pi$ into $M$, it turns out to be convenient
  (see \cite {sing}) that, for the second type of solutions, the metric
  signature in $M$ be $(-,+,+,+,-)$. For the first type it should be
  $(-,+,+,+,+)$.  If we define $|c_0| = g_0^2$, then the solution have the
  following form:
\bear \label{zac}
    X^0 \eql t \sqrt{C} \frac{2g_0}{t^2 - r^2 - g_0^2},
\nn
    X^l \eql \sigma^l \sqrt{C} \frac{2g_0}{t^2 - r^2 - g_0^2},
\nn
    X^4 \eql \sqrt{C}  \frac{t^2 - r^2 + g_0^2}{t^2 - r^2- g_0^2}
\ear
  for the first type and
\bear \label{otc}
    X^0 \eql  t \sqrt{|C|} \frac{2g_0}{t^2 - r^2 + g_0^2)},
\nn
    X^l \eql \sigma^l \sqrt{|C|} \frac{2g_0}{t^2 - r^2 + g_0^2},
\nn
    X^4 \eql \sqrt{|C|} \frac{t^2 - r^2 - g_0^2}{t^2 - r^2 + g_0^2}
\ear                                                           
  for the second type.

  To study the global properties of the manifold, let us study its
  boundaries. For the second type of solutions, consider a range $W_o$
  specified by the following constraints on the coordinate variables:
\beq \label {cor1} \nq                                                
   - |g_0| \leq (t{+}r) \geq |g_0|, \qquad -|g_0| \leq(t{-}r) \geq |g_0|.
\eeq
  Let us introduce the new coordinate $(\eta, \chi)$ instead of $(t, r)$:
\bear \label {cor2}
    t +r  \eql  g_0 \tanh\biggl(\frac{1}{2}(\eta+ \chi)\biggr),
\nn
    t - r \eql  g_0 \tanh\biggl(\frac{1}{2}(\eta - \chi)\biggr). 
\ear
  In the new coordinates, using a conformal transformation, the metric may be
  brought to a form exactly coinciding with the open
  anti-de Sitter space metric
\beq  \label{met}
    d s^2 = a^2 (\eta) [ d \eta^2 - (d\chi)^2 - K (\chi)d \Omega^2],
\eeq
  where
\[
  a^2 (\eta) = \frac{C}{\cosh^2 {\eta}}, \qquad K (\chi) = \sinh^2 {\chi},
\]
  and $d \Omega^2$ is the metric form of a $(p-1)$-sphere of unit radius,
  expressed in spherical coordinates.

  In a similar way, for the first type of solutions, the whole range $W_cl :
  -\infty < t < +\infty,\ -\infty < r < +\infty$ may be mapped into a part
  of the compact (closed) de Sitter space.

  To this end, we introduce new coordinates by the relations
\bear \label{corz2}
    t +r \eql g_0 \tan\biggl(\frac{1}{2}(\eta +\eta_0 + \chi)\biggr),
\nn
     t - r \eql g_0 \tan\biggr(\frac{1}{2}(\eta +\eta_0- \chi)\biggr),
\ear
  with $\eta_0 = \const$. The metric has the form (\ref{met}), where
\beq \label{zakrit}
    a^2 (\eta) = \frac{C}{\cos^2 {(\eta+\eta_0)}}, \qquad
      K (\chi) = \sin^2{\chi},
\eeq

  The functions $X^A$ are scalars with respect to the above coordinate
  transformations and may be rewritten in the new coordinates. For instance,
  for de Sitter space,
\beq                     \label{zakritx}            \nq
    X^{0} = \sqrt{C} \tan {(\eta{+}\eta_0)}, \qquad
      X^a = \frac{\sqrt{C}}{\cos {(\eta + \eta_0)}} k^a,
\eeq
   where $k^a $ are the embedding functions of a $p$-dimensional sphere.
   For dimension $p=3$, these functions are
\bearr
    k^1 = \sin {\chi} \sin {\theta} \cos {\phi},\qquad
    k^2 = \sin {\chi} \sin {\theta} \sin {\phi},
\nnn
    \label{ka} k^3 = \sin {\chi} \cos {\theta}, \qquad
    k^4 = \cos {\chi}.
\ear
   In this case, the metric form is
\beq              \nhq    \wide
    d s^2 = a^2 (\eta) [ d \eta^2 - d \chi^2 -
    \sin^2\chi (d \theta^2 + \sin^2 \theta d\phi^2],
\eeq
   which corresponds to the Robertson-Walker metric describing the Friedmann
   cosmological models.

   The solutions for anti-de Sitter space are obtained from the above
   equations if one makes there the following substitution:
\[
    \sin \chi \mapsto \sinh \chi,\quad
    \cos \chi \mapsto \cosh \chi, \quad
    \cos \eta \mapsto \cosh \eta.
\]

\subsection{Induced gravity as a result of a spontaneous
        violation of the conformal invariance}

  In addition to considering different gauge conditions, let us note that
  the field equations and constraint equations may also be studiedly the
  canonical gauge. To do so, using the substitution $X = b^{4\xi p} g$,
  without imposing the supplementary condition (\ref{C}), one can entirely
  exclude the field $\lambda_M$ in the Hamiltonian equations if the
  condition (\ref{n1}) is valid. In this case, the following equations are
  obtained:
\bearr  \label{kar}
    \ddot g =\stackrel{\omega}{\Delta} g + 2\xi g \stackrel{\omega}{R} +
    2\rho \Lambda g Z^{\rho - 1}, \qquad  Z \equiv (g, g),        
\yyy    \label{gry}
    \frac{1}{2\xi} [ (D_l g, D_k g) - B_g \omega_{lk}] + D_l D_k Z
\nnn \cm
    - \frac{1}{p} \stackrel{\omega}{\Delta} Z \omega_{lk}
    - [\stackrel{\omega}{R}_{lk} - \frac{1}{p}\stackrel{\omega}{R}
                \omega_{lk}] Z = 0,
\yyy  \label{ghy}
    (\dot g, \dot g) + B_g p + 4 \xi \stackrel{\omega}{\Delta} Z
    - 2 \xi Z \stackrel{\omega}{R} -2 \Lambda Z^{\rho} = 0,
\nnn \cm\cm
    B_g \equiv \frac{1}{p} \omega^{lk} (D_l g, D_k g),
\yyy    \label{gly} D_l
            \dot Z + \frac{1}{2\xi} (\dot g, D_l g) = 0.
\ear
  The latter two equations are equivalent to the constraint equations $H_0$,
  $H_l$. As their consequence, we obtain an equation for the function $Z$:
\beq\label{gzy}
    \ddot Z = (1 - 8 \xi) \stackrel{\omega}{\Delta}Z
     + 8 \xi Z \stackrel{\omega}{R} - \frac{1}{\xi} \Lambda Z^{\rho}- 4d B_g.
\eeq
  \eq (\ref{kali}) has the form
\beq
    D_l D_k Z - \frac{1}{p}\stackrel{\omega}{\Delta} Z \omega_{lk}
    -\biggl [ \stackrel{\omega}{R_{lk}}
    - \frac{1}{d}\stackrel{\omega}{R} \omega_{lk}\biggr] Z = 0,
\eeq

  We will seek special solutions to \eqs (\ref{kar})--(\ref{gzy}) for the
  dimension $p = 3$, when the metric $\omega_{lk}$ is determined by the
  3-dimensional part of the linear element of an open-type Robertson-Walker
  space-time. We seek solutions in the form
\beq \label{gnc}
    g_0 = u_{0}(\eta), \qquad g^a = u (\eta) k^a (\sigma^i),
    \quad a = 1,2,3,4,
\eeq
  and, doing so, we do not require that the full embedding conditions
  (\ref{kal}) should hold. Then, for the functions $u_{0}(\eta)$ and $u(\eta)$
  we obtain
\bearr \label{gnc1}
    \dot u^2 + 4k u^2 - 8k\Lambda  \int (u^2 - k u_0^2)u du - 2H = 0,
\nnn \inch
        H = \const,                        
\yyy \label{gnc2}
    \dot u_{0}^2 +  k u_0^2 - 8k\Lambda  \int (u^2 -k u_{0}^2) u_{0}
            du_{0} + 2H = 0.           
\ear
  The last two equation with respect to the variables $u_{0}(\eta)$ and
  $u(\eta)$) may be considered as a dynamic system with the potential energy
\[  \nq
    U(u, u_0) = -k \Lambda u^2 (u^2 +2 u_{0}^2)+\Lambda u_{0}^4
            +k(2 u^2 - u_0^2 /2)
\]
  and zero total energy. Integrating by parts and summing, we can obtain
\[
    \dot u^2 + \dot u_{0}^2  + 2U = 0.
\]
  A further study shows that the previously found solution, describing an
  open de Sitter space, is a stable exceptional solution to \eqs
  (\ref{gnc1})--(\ref{gnc2}). In the present formulation, the following
  solution corresponds to the one obtained above:
\beq   \nq \label{gnc3}
    u = \frac{r^2}{(\cosh {\eta})^2},\quad\
    u_0 = \frac{r^2 \tanh {\eta}}{\cosh \eta}, \quad\
    r^2 = \sqrt{\frac{3}{2|\Lambda|}}.                     
\eeq

   Using the terminology of the qualitative theory of differential
   equations, the singular point $u = 0$, $u_0 = 0$ is unstable.
   There are no other static points.  Meanwhile, the solution (\ref{gnc3}),
   being a separatrix in the phase space of the variables $u_{0}(\eta)$ and
   $u(\eta)$, minimizes the total energy.

   From this point of view, it is of interest to invoke the Higgs mechanism
   to obtain the constraints (\ref{gi}). The fields $X^A$, being coordinates
   of the space $M$, may play the role of Higgs' fields in Grand Unification
   models. On the other hand, from the viewpoint of the Hamiltonian
   formalism considered above, solutions with a broken symmetry may be
   treated as a particular choice of the gauge.

\subsection{The hierarchy of coupling constants}

   As has been already noted above, we here obtain equations similar to the
   Einstein equations with an effective gravitational constant, see
  (\ref{kap}). Indeed, in case $(X,X)=C$ and if
\beq \label{gti}
    g_{\mu \nu} =(\nabla_{\mu} X, \nabla_{\nu} X), \cm
        \mu,\nu =\overline{0,p},
\eeq
  \eqs (\ref{T3}) take the form
\beq \label{TE3}
      G_{\alpha\beta} =8 \pi G_e T^{e}_{\alpha\beta}
                     + \Lambda_e  g_{\alpha\beta},
\eeq
  where $G_e$ is given by \eq (\ref{kap}), while the cosmological constant is
\beq \label{lamcos}
        \Lambda_e=- \frac{1}{2\xi C} (-1+\Lambda C^2).
\eeq
   From the solution (\ref{cosm}) and (\ref{lamcos}), we find that
\beq \label{coslam}
    \Lambda = \frac {3}{2C^2},  \cm  \Lambda_e = \frac {3}{C}.
\eeq

   In a closed model, the constant $C$ satisfies the equation
\beq \label{constc}
    -(X^0)^2+(X^1)^2+(X^2)^2+(X^3)^2+(X^4)^2=C.          
\eeq
   Then $\sqrt{C}$ characterizes the size of the observed part of the
   Universe. In the solution (\ref{zakrit}) (for $\eta_0=\pi/2$), we pass
   over to the proper time $t$ and obtain:
\bear \label{massht}
    a(t) \eql \sqrt{C}\cosh(t/\sqrt{C}),
\nn
     H \al \equiv\al \frac{\dot{a}}{a}=\sqrt{C}\tanh(t/\sqrt{C}).
\ear
  Suppose that the Hubble ``constant'' $H\sim (3 \cdot 10^{17})^{-1}
  c^{-1}$ (in the Planck units, $\hbar = 1$ and $c = 1$) and that our epoch
  corresponds to the time $t\simeq \sqrt{C}$, then we obtain the calue of
  $C$: $\sqrt{C}\simeq 7.2 \cdot 10^{27}$ cm $\sim 10^{28}$ cm.
  Substituting this value into (\ref{coslam}), we find
  $\Lambda_e\sim 10^{-56}\ {\rm cm}^{-2}$, or, the same in energy units,
  $\Lambda_e \sim 10^{-46}\ GeV^4$. This result confirms the existence of a
  nonzero cosmological constant $\Lambda$, which is also in agreement with
  the observational data, see, e.g., Ref.\,\cite{starob}.

  Equating the expression (\ref{kap}) to $1/{M_p^2} \sim
  10^{-66}\ {\rm cm}^2$, we find that $w_0\sim 4\cdot 10^{-10}cm^4$.
  The parameter $w_0$ also corresponds to distances of the order
  $l_w = \sqrt[4]{w_0}\sim 0.05$ mm. To explain the nature of the emerging
  scale, one can invoke the Randall-Sundrum conjecture \cite{randal1}, where
  the existence of extra dimensions ($n>4$) is supposed, with a sufficiently
  small size ($l < 0.2$ mm) for being in agreement with the experimental
  data.

  In conclusion, let us note that if we consider the action obtained from
  (\ref{ns1}) by adding to it the Einstein term $(1/G) R$, which violates
  the conformal invariance of the equations, then there emerges the
  effective gravitational ``constant'' $G_e = wG (w+2\xi G (X, X))^{-1} $ .
  As is shown in Ref.\,\cite {zar1}, this leads to an instability of
  cosmological solutions for $G_e \to 0 $. This result is one of the
  arguments in favour of consideration of an intially conformally invariant
  theory of gravity; this invariance will then be probably violated due to
  quantum effects.


\small
\begin{thebibliography}{20} \itemsep 2pt

\bibitem {grin}
     Œ.B. Green, J.H. Schwarz and E. Witten, ``Superstring theory'',
     Cambridge University Press, 1987.

\bibitem{brn}
     L. Brink and Œ. Henneaux,
     ``Principles of String Theory'', Plenum Press, NY--London, 1988.

\bibitem {vit}
    B. d¥ Wit,  ``Introduction to Supergravity'',  M., Mir, 1985.

\bibitem {zar}
     F.Sh. Zaripov, Proc. Int. School-Seminar ``Foundation of the Theory of
     Gravity and Cosmology'', Odessa, 1995, p. 35.

\bibitem {ket}
     S.V. Ketov, ``Introduction to Quantum Theory of Strings and
     Superstrings'', Nauka, Novosibirsk, 1990 (in Russian).

\bibitem {randal1}
      L. Randall and R. Sundrum, \PRL {83} 3370 (1999); hep-ph/9905221.

 \bibitem {randal2}
      L. Randall and R. Sundrum, \PRL {83} 4690 (1999).

\bibitem{barvinski}
      A.O. Barvinsky, hep-th/9906064; {\it Phys. Uspekhi\/} {\bf 175} (6),
      569--601  (2005).

\bibitem{rub}
      V.A. Rubakov, {\it Phys. Uspekhi\/} {\bf 173} (2), 219--226 (2003).

\bibitem{god}
      P. Goddard and C.B. Thorn, ``Compatibility of the dual pomeron
      with unitarity and the absence of ghosts in the dual resonance
      model'', \PLB  {7} 535 (1972).

\bibitem{duf}
      M.J. Duff, ``Supermembranes: the First Fifteen Weeks'',
      {\it preprint\/} CERN-TH-4797, Geneva, 1987.

\bibitem {lin}
      U. Lindstrom and M. Rocek, ``A super-Weyl-invariant spinning
      membrane'',  {\it preprint\/} ITP-SB-61, NY, 1988.

\bibitem {milna}
      E.A. Milne, {\it Nature (London)\/} {\bf 130}, 9 (1933).

\bibitem {zar1}
      F.Sh. Zaripov, ``On the stability of the Friedmann universe with
      a conformally invariant scalar field with self-interaction'', {\it
    in:\/} ``Gravitation and Relativity'', Kazan State University
    Press, Kazan, 1986, No. 23, pp. 62--75.

\bibitem{tag}
      E.A. Tagirov and N.A. Chernikov, ``Quantum theory of a scalar field
      in de Sitter space'', {\it Ann. Inst. Hent\'\i \ Poincar\'e\/} {\bf 9},
      109 (1968).

\bibitem{tor}
      C.W. Misner, K.S. Thorne and J.A. Wheeler,  ``Gravitation'',
      W.H. Freeman and Co., San Francisco, 1973.

\bibitem{tut}
       D.M. Gitman and I.V. Tiutin, ``Canonical Quantization of Fields
       with Constraints'',  M., Mir, 1986.

\bibitem{sing}
       J.L. Synge, ``Relativity: the General Theory'', North Holland,
       Amsterdam, 1960.

\bibitem{starob}
       V. Sahni and A.A. Starobinsky, {\it Int. J. Mod. Phys.\/} {\bf D 9},
         373 (2000); astro-ph/9904398.

\end{thebibliography}
\end{document}